\documentclass[twoside]{LCWS11}
\usepackage[latin1]{inputenc}
\usepackage[dvips]{graphicx,epsfig,color}
\usepackage{wrapfig,rotating}
\usepackage{amssymb,amsmath,array}

\usepackage{cancel}

\begin{document}
\title{
Multi-tau lepton signatures in leptophilic two Higgs doublet
model at the ILC} 
\author{Shinya Kanemura$^1$, Koji Tsumura$^2$ and Hiroshi Yokoya$^3$
\vspace{.3cm}\\
1- 
The University of Toyama - Department of Physics\\ 
Toyama 930-8555 - Japan
\vspace{.1cm}\\
2- 
National Taiwan University - Department of Physics and Center for Theoretical Sciences\\
Taipei 10617 - Taiwan
\vspace{.1cm}\\
3-
National Taiwan University - National Center for Theoretical Sciences\\
Taipei 10617, Taiwan
\\
}

\maketitle

\begin{abstract}
%
We study the feasibility of the Type-X two Higgs doublet model (THDM-X) at
collider experiments.
In the THDM-X, new Higgs bosons mostly decay into tau leptons 
in the wide range of the parameter space. 
Such scalar bosons are less constrained by current experimental data, 
because of the suppressed quark Yukawa interactions. 
We discuss a search strategy of the THDM-X with multi-tau lepton final states 
at International linear collider and Large Hadron Collider.
By using the collinear approximation, we show that a four tau lepton signature 
$(e^+e^-\to HA \to 4\tau)$ can be a clean signal. 
\end{abstract}

\section{Introduction}

The Higgs sector is unknown, because no Higgs boson has been discovered yet. 
In the Standard Model (SM) for elementary particles, only one scalar iso-doublet 
field is introduced to break the electroweak gauge symmetry spontaneously. 
However, as the various models beyond the SM predict the extended Higgs sector, 
there is a possibility of non-minimal Higgs sectors.

The non-minimal Higgs sectors suffer from the constraints from 
the rho parameter and the flavor changing neutral current (FCNC) in general.
In the SM, these constraints are automatically satisfied. 
It is known that in the Higgs sector with only doublets, the rho parameter
is predicted to be unity at the tree level.
Therefore, two Higgs doublet models (THDMs) would be a simplest viable
extension of the SM.
However, in the THDM the most general Yukawa interaction predicts FCNC
at the tree level, because both the doublet couples to a fermion so that
the mass matrix and the Yukawa matrix cannot be diagonalized simultaneously.
In order to avoid this, a discrete symmetry may be introduced under which
different properties are assigned to each scalar doublet~\cite{Ref:GW}.
Under this symmetry, each fermion couples with only one 
scalar doublet, and hence there is no FCNC at the tree level even in the THDM. 

There are four types of Yukawa interactions depending on the
$Z_2$-charge assignments; i.e., Type-I, II, X and Y. 
An interesting possibility would be the Type-X THDM, 
where one Higgs doublet couples with quarks and the other does with
leptons~\cite{Ref:Barger,Ref:AKTY,Ref:TypeX}.
The Type-X THDM can appear in the Higgs sector of a gauged
extension of the Type-III seesaw model~\cite{Ref:GaugedTypeIII}, 
the model of the three-loop neutrino mass with electroweak baryogenesis~\cite{Ref:AKS} 
and a model for the positron cosmic ray anomaly~\cite{Ref:Hall}.

\section{The Type-X two Higgs doublet model}

The Higgs potential of the THDM is defined as~\cite{Ref:HHG,Ref:Djouadi2}
\begin{align}
{\mathcal V}^\text{THDM} 
&= +m_1^2\Phi_1^\dag\Phi_1+m_2^2\Phi_2^\dag\Phi_2
-m_3^2\left(\Phi_1^\dag\Phi_2+\Phi_2^\dag\Phi_1\right)
+\frac{\lambda_1}2(\Phi_1^\dag\Phi_1)^2
+\frac{\lambda_2}2(\Phi_2^\dag\Phi_2)^2\nonumber \\
&\qquad+\lambda_3(\Phi_1^\dag\Phi_1)(\Phi_2^\dag\Phi_2)
+\lambda_4(\Phi_1^\dag\Phi_2)(\Phi_2^\dag\Phi_1)
+\frac{\lambda_5}2\left[(\Phi_1^\dag\Phi_2)^2
+(\Phi_2^\dag\Phi_1)^2\right], \label{Eq:HiggsPot}
\end{align}
where $\Phi_i(i=1,2)$ are the Higgs doublets with hypercharge $Y=1/2$. 
A softly broken $Z_2$ symmetry is imposed in the model to forbid FCNC at
the tree level, under which the Higgs doublets are transformed as
$\Phi_1\to +\Phi_1$ and $\Phi_2 \to -\Phi_2$~\cite{Ref:GW}. 
The soft-breaking parameter $m_3^2$ and the coupling constant
$\lambda_5$ are complex in general. 
We here take them to be real assuming the CP invariant Higgs sector. 

The Higgs doublets can be written in terms of the component fields as
\begin{align}
\Phi_i=\begin{pmatrix}i\,\omega_i^+\\\frac1{\sqrt2}(v_i+h_i-i\,z_i)
\end{pmatrix},
\end{align}
where the vacuum expectation values (VEVs) $v_1$ and $v_2$ satisfy 
$\sqrt{v_1^2+v_2^2}=v \simeq 246$ GeV and $\tan \beta \equiv v_2/v_1$.
The mass eigenstates are obtained by rotating the component fields as
\begin{align}
\begin{pmatrix}h_1\\h_2\end{pmatrix}=\text{R}(\alpha)
\begin{pmatrix}H\\h\end{pmatrix},\quad
\begin{pmatrix}z_1\\z_2\end{pmatrix}=\text{R}(\beta)
\begin{pmatrix}z\\A\end{pmatrix},\quad
\begin{pmatrix}\omega_1^+\\\omega_2^+\end{pmatrix}=\text{R}(\beta)
\begin{pmatrix}\omega^+\\H^+\end{pmatrix},
\end{align}
where $\omega^\pm$ and $z$ are the Nambu-Goldstone bosons, $h$, $H$, $A$
and $H^\pm$ are respectively two CP-even, one CP-odd and charged Higgs
bosons, and
\begin{align}
\text{R}(\theta)=\begin{pmatrix}\cos\theta&-\sin\theta \\
\sin\theta&\cos\theta\end{pmatrix}.
\end{align}
The eight parameters $m_1^2$--$m_3^2$ and
$\lambda_1$--$\lambda_5$ are replaced by the VEV $v$, the mixing angles
$\alpha$ and $\tan\beta$, the Higgs boson masses
$m_h^{},m_H^{},m_A^{}$ and $m_{H^\pm}^{}$, and the soft $Z_2$ breaking
parameter $M^2=m_3^2/(\cos\beta\sin\beta)$.
The coupling constants of the CP-even Higgs bosons with weak gauge bosons
$h VV$ and $H VV (V=W,Z)$ are proportional to $\sin(\beta-\alpha)$ and
$\cos(\beta-\alpha)$, respectively.
When $\sin(\beta-\alpha) =1$, only $h$ couples to the gauge bosons
while $H$ decouples. 
We concentrate on this limit (the SM-like limit) where $h$ behaves 
as the SM Higgs boson~\cite{Ref:GunionHaber,Ref:KOSY}. 

Imposing the transformation under the $Z_2$ parity for leptons and quarks as,
$u_R\to -u_R$, $d_R\to -d_R$, $\ell_R\to +\ell_R$,
$Q_L\to +Q_L$ and $L_L\to +L_L$, we could write down the Type-X Yukawa
interaction~\cite{Ref:Barger,Ref:AKTY};
\begin{align}
{\mathcal L}_\text{yukawa}^\text{Type-X} =
&-{\overline Q}_LY_u\widetilde{\Phi}_2u_R^{}
-{\overline Q}_LY_d\Phi_2d_R^{}
-{\overline L}_LY_\ell\Phi_1 \ell_R^{}+\text{H.c.}
\end{align}

In the Type-X THDM, more than $99\%$ of $H$ and $A$ decay into 
pairs of tau leptons for $\tan\beta\gtrsim 3$ in the SM-like limit; 
$\sin(\beta-\alpha)=1$~\cite{Ref:AKTY}. 
The neutral Higgs bosons would be produced in pair by $q \bar q \to Z^*
\to HA$ process at the Large Hadron Collider (LHC) and by $e^+e^- \to Z^*
\to HA$ process at the International Linear Collider (ILC). 
These Higgs bosons predominantly decay into a four-$\tau$ state, 
$HA\to (\tau^+\tau^-)(\tau^+\tau^-)$, which is the characteristic signal 
of the Type-X THDM. 
There would be a clear signature in the dimuon channel from the direct decay 
of the Higgs bosons, $HA\to (\mu^+\mu^-)(\tau^+\tau^-)$. 
Although the number of events is only $2(m_\mu/m_\tau)^2 \sim 0.7 \%$ 
of the four tau lepton channel, this channel would be important to measure 
the mass of the Higgs bosons at the LHC. 

Experimental constraints on masses of the neutral Higgs bosons $H$, $A$ in 
THDMs depend on the type of the Yukawa interaction. 
In the Type-II THDM with large $\tan\beta$, stronger mass bounds can be obtained
from these production processes at the Tevatron and the 
LHC~\cite{Ref:SUSYHiggsTeV,SUSYHiggsLHC}. 
However, if the Yukawa interaction is the lepton specific
which is realized in the wide parameter space in the Type-X THDM, 
these Higgs bosons are less constrained 
due to the relatively weak Yukawa interaction with quarks. 
The search for such Higgs bosons at the LEP experiments is found in
Ref.~\cite{Ref:LEP4tau}. 

\section{Simulation study}

\subsection{The collinear approximation}

In our analysis, we use the collinear approximation to calculate the
four momenta of the tau leptons\cite{Ref:HRZ}. 
If tau leptons are energetic, the missing momentum from its decay would
be along the direction of the charged track (either a charged hadron
(hadrons) or a charged lepton), ${\vec p}^{\; miss} \simeq c\, {\vec
p}^{\; \tau_j}$, where ${\vec p}^{\; miss}$, ${\vec p}^{\; \tau_j}$ are
the momenta of the neutrino and the charged track, respectively.
The proportionality constant $c$ can be determined by fixing ${\vec
p}^{\; miss}$. 
Accordingly, the momentum of the decaying tau lepton can be
approximately reconstructed as ${\vec p}^{\; \tau} \simeq (1+ c)\, {\vec
p}^{\; \tau_j} \equiv z^{-1}\, {\vec p}^{\; \tau_j}$, where $z$ is the
momentum fraction of the charged track from the parent tau lepton.

At hadron colliders, the transverse components of the missing momentum
$\vec{\cancel{p}}_T^{}$ can be measured as the negative sum of the visible momenta. 
Assuming that the missing transverse momentum of the event is accounted
solely by the missing particles in the decays of tau leptons, and applying
the collinear approximation for two tau leptons, the missing transverse
momentum can be expressed by the momenta of charged tracks, as
$\vec{\cancel{p}}_T^{} \simeq {\vec p}^{\; miss_1}_T + {\vec p}^{\;
miss_2}_T \simeq c_1\, {\vec p}^{\; \tau_{j1}}_T + c_2\, {\vec p}^{\;
\tau_{j2}}_T$.
Unknown parameters $c_1$ and $c_2$ are determined by solving simultaneous
equations.
Using the resulting values of $z_1$ and $z_2$, 
the invariant mass of the tau lepton pair is related with that of
the tau-jet pair as $M_{\tau_h\tau_h}^2 \simeq z_1 z_2
M_{\tau\tau}^2$.

At $e^+e^-$ colliders, neutral Higgs boson pair can be produced 
via $e^+e^- \to H A$, and the four momenta of the four tau leptons 
are completely solved~\cite{Ref:LEP4tau,Ref:LEPdouble}.

\subsection{The $2\mu2\tau_h$ channel at the LHC}
The signal events are generated by using {\tt PYTHIA}~\cite{Sjostrand:2006za}, 
where the decay of tau leptons is simulated by using {\tt TAUOLA}~\cite{Jadach:1993hs}.
Initial-state-radiation (ISR) and final-state-radiation (FSR) effects
are included. 
We choose the collision energy to be $14$~TeV, and use the {\tt CTEQ6L} 
parton distribution functions~\cite{Pumplin:2002vw}.
We set the masses of extra Higgs bosons to $m_H^{}=130$ GeV, $m_A^{}=170$ GeV.
The total cross section for $pp\to HA$ is estimated to be
$53$~fb at the tree level~\cite{Ref:AKTY}. 
For the LHC study, background events for $VV$ ($= ZZ$, $ZW$ and $WW$), 
$t\bar t$ processes where the weak bosons decay leptonically and hadronically, 
and  $Z+$jets processes followed by leptonic decays of weak
bosons are generated by {\tt PYTHIA}, where the decays of tau leptons
are also handled by {\tt TAUOLA}.
The total cross sections for these processes are given as $108$~pb,
$493$~pb and $30$~nb, respectively for $VV$, $t\bar t$ 
and $Z$+jets production processes by {\tt PYTHIA}. 

We identify the tau-jet candidates by
the following criteria; 
 a jet with $p_T^{}\ge 10$~GeV and $|\eta|\le2.5$ which contains $1$ or $3$ 
 charged hadrons in a small cone ($R=0.15$) centered at the jet momentum 
 direction with the transverse energy deposit to this small cone more 
 than $95 \%$ of the jet.

In order to evaluate the signal significance, we use the significance
estimator $S$ defined as~\cite{Ref:CMS-TDR} 
\begin{align} 
S &= \sqrt{2 \bigl[ (s+b) \ln(1+s/b)-s \bigr]},
\end{align} 
where $s$ and $b$ represent the numbers of signal and background
events, respectively.
The significance $S$ is proportional to the square root of the
integrated luminosity. 

\begin{table}[tb]
 \begin{center}
 \begin{tabular}{|c||c||c|c|c||c|}
  \hline $2\mu2\tau_h$ event analysis & $HA$ & 
  $VV$ & $t\bar t$ & $Z+$jets & $S$ (100~fb$^{-1}$)\\ \hline \hline
  Pre-selection
  & 87.3 & 350.6 & 767.9 & 28785.9 & 0.50 \\
  $p_T^{\tau_h} > 40$~GeV
  & 45.9 &  96.5 & 154.1 &  4397.3 & 0.67 \\
  $\cancel{E}_T^{} > 30$~GeV
  & 37.6 &  49.9 & 134.9 &    37.1 & 2.5 \\
  $H_T^\text{lep} > 250$~GeV
  & 20.6 &  16.9 &  48.5 &     0.  & 2.4 \\
  $H_T^\text{jet} < 50$~GeV
  & 14.1 &  11.3 &   4.1 &     0.  & 3.2 \\
  \hline
  $0 \le z_{1,2} \le 1$
  &  3.5 &   7.9 &   0.6 &     0.  & 1.1 \\
  $(m_Z^{})_{\mu\mu} \pm 10$~GeV
  &  3.3 &   1.0 &   0.5 &     0.  & 2.1 \\
  $(m_Z^{})_{\tau\tau} \pm 20$~GeV
  &  3.1 &   0.2 &   0.5 &     0.  & 2.6 \\
  \hline
  \end{tabular}
 \end{center}
\caption{Table for background reductions in the $2\mu2\tau_h$ channel.
 Listed are the expected number of events for the integrated luminosity
 of 100~fb$^{-1}$ at the LHC with $\sqrt{s}=14$~TeV.
 }
\label{Tab:LHC_2T2M}
\end{table}

The results of the signal/background reduction are summarized 
at each step in TABLE~\ref{Tab:LHC_2T2M}.
We show the expected numbers of events for the integrated luminosity of
$L=100$~fb$^{-1}$ for each process.
The signal events consist of the hadronic decay of tau leptons with 
the primary muons from the Higgs bosons as well as the secondary muons 
from the tau leptonic decay. 
Background events from the $Z+$jets process contain two mis-identified
tau-jets from the ISR jets, with a muon pair which comes from
the $Z/\gamma^*\to \mu^+\mu^-$ decay. 
Therefore, $Z+$jets background events tend to have small $\cancel{E}_T$, and
the cut on $\cancel{E}_T$ is expected to reduce the $Z+$jets background
significantly.
The cut on $H_T^\text{lep}$ can reduce the $VV$ and $Z+$jets backgrounds
significantly. 
The background contribution from the $t\bar{t}$ events can be reduced by
using the cut on $H_T^\text{jet}$, because the $t\bar t$
events tend to contain many jets due to the $b$ quark fragmentation and
ISR/FSR, even though two of them are mis-identified as tau-jets. 
Furthermore, the events which contain $Z\to\mu^+\mu^-$ can be reduced 
by rejecting the events with the invariant mass of the muon pair close to
$m_Z^{}$.

The largest significance can be obtained after the $m_Z^{}$-window cut 
of the dimuon, where the number of the signal events is expected to be 
about $14$ while that of background events is about $19$ giving 
$s/b\sim 1$ and $S \sim 3.2$ for $L=100$~fb$^{-1}$~\footnote{
Further optimization of the kinematical cuts and the analysis in other 
decay channels, $3\mu1\tau_h$, etc., have been studied in Ref.~\cite{Ref:KTY}
}.
For the $S=5$ discovery of the signal, we found that 
the integrated luminosity of about $300$ fb$^{-1}$ is required.

By using the collinear approximation, we can reconstruct the tau lepton 
momenta and extract the events with the primary muons from the Higgs boson decay.
For this $2\mu2\tau$ signal, the $VV$ background can be further reduced 
by the cut on the $m_Z^{}$-window for the reconstructed $M_{\tau\tau}$. 
Even if we focus on the signal only from the $HA\to 2\mu2\tau$ mode, 
the signal can be tested almost at the same level as the dimuon invariant mass analysis. 
The extraction of this mode using the collinear approximation would be useful 
to determine the mass of Higgs bosons accurately. 

\subsection{The $4\tau_h$ channel at the ILC}
The neutral Higgs bosons can be pair produced via the $e^+e^-\to HA$ 
process, and their decay produces four tau lepton final states dominantly. 
At $e^+e^-$ colliders, the four momenta of the four tau leptons can be
solved by applying the collinear approximation to all the four decay
products of the tau leptons~\cite{Ref:LEP4tau,Ref:LEPdouble}, because
the missing four momentum can be reconstructed by the energy momentum
conservation. 
In our analysis, we choose the collision energy to be $500$~GeV and 
the signal cross section is $30$~fb. 
The cross sections of the background processes are given as $8300$~fb 
and $580$~pb respectively for $VV$ and $t\bar t$ . 

\begin{table}[tb]
 \begin{center}
 \begin{tabular}{|c||c||c|c||c|}
  \hline $4\tau_h$ event analysis & $HA$ & 
  $VV$ & $t\bar t$ & $S$ (100~fb$^{-1}$)\\ \hline \hline
  Pre-selection
  & 300. & 10.6 & 1.2 & 38. \\
  $0 \le z_{1-4} \le 1$
  & 251. &  6.2 & 0.1 & 38. \\
  $(m_Z^{})_{\tau\tau} \pm 20$~GeV
  & 238. &  1.8 &  0. & 43. \\
  \hline
  \end{tabular}
 \end{center}
\caption{Table for background reductions in the $4\tau_h$ channel.
 Listed are the expected number of events for the integrated luminosity
 of 100~fb$^{-1}$ at the ILC with $\sqrt{s}=500$~GeV.
 }
\label{Tab:ILC_4T}
\end{table}

The results of the signal/background reduction are summarized in TABLE~\ref{Tab:ILC_4T}.
The expected numbers of events are normalized for the integrated luminosity of
$L=100$~fb$^{-1}$ for each process. 
We here focus on the hadronic decay mode of all tau leptons. 
In general, the mixture of the hadronic and the leptonic decay modes can be analysed. 
And the significance can be improved by combining the all channels.
In order to construct the invariant mass of the tau lepton pair from four tau leptons, 
we choose the combination of the opposite signed tau leptons which gives the highest 
$p_T^{}$ pair. 

The signal events are dominant even at the pre-selection level. 
The statistical significance can be further optimized by using the kinematical cuts 
giving the much better $s/b$ ratio. 
In order to test the signal with $S=5$, we only need the integrated luminosity of 
about $5$ fb$^{-1}$ where we only use the $4\tau_h$ channel.

\section{Summary and Conclusion}

We have presented the simulation study of the tau lepton specific Higgs bosons 
at the LHC and the ILC. 
In the THDM-X with the SM-like limit, the additional Higgs bosons can be 
the tau lepton specific. 
Such scalar bosons can be pair produced by the gauge interaction at 
the LHC and the ILC, and mainly decay into tau leptons in the wide 
range of the parameter space.
By using the collinear approximation, we show that multi-tau lepton final 
state $HA \to 2\mu2\tau$ at the LHC and $HA\to 4\tau$ at the ILC 
can be a clean signal. 
The tau lepton specific Higgs boson can be tested at the LHC with 
about $300$ fb$^{-1}$ of the integrated luminosity for $S=5$. 
Although the huge integrated luminosity is required, the precise mass 
determination is possible by extracting the primary muon from the 
Higgs boson decay in the $2\mu2\tau_h$ channel.
The search potential of the ILC with $4\tau_h$ channel is about $70$ times 
better than that of the LHC with the $2\mu2\tau_h$ channel in the sense 
of the integrated luminosity.
Since the $4\tau_h$ channel can be fully reconstructed by the collinear 
approximation, the mass of Higgs bosons can also be measured.

\begin{footnotesize}

\section*{Acknowledgments}
The work of S.K.\ was supported in part by Grant-in-Aid for Scientific
Research, Japan Society for the Promotion of Science (JSPS),
Nos.~22244031 and 23104006.
The work of K.T.\ was supported in part by the National Science Council
of Taiwan under Grant No.~NSC 100-2811-M-002-090.
The work of H.Y.\ was supported in part by the National Science Council
of Taiwan under Grant No.~NSC 100-2119-M-002-001.


\end{footnotesize}


\end{document}